\documentstyle[11pt]{article}
\def\be{\begin{equation}}
\def\ee{\end{equation}}
\def\al{\alpha}
\def\bt{\beta}
\def\ld{\lambda}

\def\gm{\gamma}

\def\sg{\sigma}

\begin{document}
\thispagestyle{empty}
\setcounter{page}0
%\vspace*{1 cm}
\begin{flushright}
%{\normalsize CERN-PH-TH/2004-080}
CERN-PH-TH/2004-080
\end{flushright} 
~\vfill
\begin{center}
{\Large \bf  Predictions for energy distribution and polarization\\ 
of the positron from the polarized muon decay}\\

\vspace{.5cm}
{\large M. V. Chizhov}
\vspace{.5cm}

{\em
Theory Division, Department of Physics, CERN,\\ 
CH-1211 Geneva 23, Switzerland\\
and\\
Centre of Space Research and Technologies, Faculty of Physics,\\
University of Sofia, 1164 Sofia, Bulgaria
}

\end{center}
\vfill

\begin{abstract}
The pure lepton decay of the polarized muon is considered,  
accounting for  a new tensor interaction which is outside of the Michel local
interactions. This interaction leads to new energy distribution and
polarization of the final charged lepton. The presence of such a type of
interaction is strongly required for the description of  the latest
experimental results on the weak radiative pion decay in the full 
kinematic region. Assuming quark--lepton universality, predictions 
for a deviation from the Standard Model are made using only one new
parameter. They do not contradict  the present experimental data
and can be verified further in the on-going experiments at PSI and
TRIUMF, at least to the  level of 3 standard deviations. 
\end{abstract}

\vfill

\newpage

\section{Introduction}
            
At present only one type of vector particles is known -- the gauge one. 
The quantum field theory of the gauge interactions has been
formulated~\cite{YM} and it has been proved to be unitary and
renormalizable~\cite{Faddeev}. The Higgs mechanism of acquiring
mass~\cite{Higgs} shows the way of application of this theory to
phenomenology~\cite{SM}. The gauge particles are described by
four-vector fields $A_\al$, which are transformed under the Lorentz 
group as a real representation (1/2,1/2). 
This {\em chirally neutral} representation can be constructed as a product 
of fundamental {\em chiral} spinor representations (1/2,0) and (0,1/2). 
The latter correspond to left-handed $\psi_L=\frac{1}{2}(1-\gm^5)\psi$ 
and right-handed $\psi_R=\frac{1}{2}(1+\gm^5)\psi$ two-component Weyl 
spinors, which are related by the discrete $P$-transformation of the 
spatial reflection or by the charge conjugation $C$. So, $CP$ is
an underlying symmetry of the Lorentz group. Moreover, the main feature 
of the Lorentz group as a relativistic generalization of Galileo's invariance
is the natural introduction of two non-equivalent {\it left-handed} and 
{\it right-handed} Weyl spinors, which describe different particles
and explicitly lead to $P$ violation in nature.

At small transfer momenta, an exchange of the massive gauge vector 
particles leads effectively to contact four-fermion vector 
interaction. It is the well known Fermi interaction for the weak 
processes, which in the case of the muon decay reads
\be
{\cal L}_V=-\frac{4G_F}{\sqrt{2}}~
\left(\bar{\mu}_L\gm_\al\nu_\mu\right)
\left(\bar{\nu}_e\gm_\al e_L\right) + {\rm h.c.},
\label{vector}
\ee
with the Fermi coupling constant given by $G_F/\sqrt{2}=g^2/8M^2_W$, where
$g$ is the gauge coupling constant and $M_W$ is the mass of the gauge
weak boson. This interaction preserves the helicities of the incoming
and the outgoing fermions.

The Higgs particles are needed for the symmetry breaking in the
Standard Model (SM). They are described by the scalar fields $H$, which
are transformed under a spinless (0,0) representation of the Lorentz
group. In an extension of the SM with more than one Higgs doublet, 
new charged scalar particles can also contribute to the muon decay
\be
{\cal L}_S=-g_{RR}^S~\frac{4G_F}{\sqrt{2}}~
\left(\bar{\mu}_R\nu_\mu\right)
\left(\bar{\nu}_e e_R\right) + {\rm h.c.},
\label{scalar}
\ee
with the negative dimensionless coupling constant $g_{RR}^S$ given by
$g_{RR}^S G_F/\sqrt{2}=-h^2/8M^2_H$, where $h$ is the Yukawa
coupling constant of the lepton currents to the charged Higgs
boson with  mass $M_H$.
The dimensionless coupling constant $g_{RR}^S$ is introduced in accordance 
to the notation of ref.~\cite{Fetscher} and determines the strength
of the new interactions relative to the ordinary weak interactions
(\ref{vector}), governed by the Fermi coupling constant $G_F$.

The scalar interaction (\ref{scalar}) includes a coupling between right-handed 
charged leptons and left-handed neutrinos, and therefore it 
does not preserve the helicities of fermions.
In general other scalar interactions with right-handed neutrinos can
be written down, but since the neutrino masses are very tiny 
they do not interfere with the standard weak interaction~(\ref{vector}) 
and their contributions are negligibly small.

It should be noted that the Lorentz group provides the possibility 
to construct {\em inequivalent chiral} representations (1,0) and 
(0,1), which also correspond to particles with spin~1. They can be
constructed if one uses only the product either of the left-handed
(1/2,0) or of the right-handed (0,1/2) fundamental spinors.
Such particles are described by the antisymmetric second-rank 
tensor fields $T_{\mu\nu}$ and they interact with tensor currents. 
An example of the presence of the new kind of {\em chiral particles} 
in nature is the existence of the axial-vector hadron resonance 
$b_1(1235)$, which has only anomalous tensor interactions with quarks.
The introduction of such excitations into the Nambu--Jona-Lasinio quark model 
leads to a successful description of the dynamical properties of spin-1 
mesons~\cite{NJL} and to hints that the analogous chiral particles 
could exist at the electroweak scale as well. An exchange of such spin-1
massive chiral bosons can effectively lead to new four-fermion
tensor interactions.

Four-fermion local tensor interactions have been introduced for 
the muon decay in ref.~\cite{Michel}. However, all of them include the
right-handed neutrinos and they do not interfere with the standard
weak interaction (\ref{vector}). The local tensor interaction involving 
both left-handed neutrinos cannot be written down owing
to the identity $\left(\bar{\mu}_R\sg_{\al\bt}\nu_\mu\right)
\left(\bar{\nu}_e\sg_{\al\bt}e_R\right)\equiv 0$. 
However, if we assume that an effective tensor interaction
arises from an exchange of the new chiral bosons, then its Lorentz
structure should reflect the Lorentz structure of the propagator for such
particles. It has been shown~\cite{MPL}, that besides the local
tensor interactions, the new non-local momentum-transfer-dependent
tensor interactions should be introduced. For the muon decay this leads
to a new tensor interaction that includes only left-handed neutrinos
\be
{\cal L}_T=-g_{RR}^T~\sqrt{2}G_F~
\left(\bar{\mu}_R\sg_{\al\ld}\nu_\mu\right)\frac{4q_\al q_\bt}{q^2}
\left(\bar{\nu}_e\sg_{\bt\ld}e_R\right) + {\rm h.c.},
\label{tensor}
\ee
where $g_{RR}^T$ is the new positive tensor coupling constant.
Such type of interaction can interfere with the standard weak interaction
(\ref{vector}) and it is more sensitive to experimental detection.

Moreover, the recent results of the PIBETA experiment on the radiative 
pion decay~\cite{PIBETA} strongly show evidence for the
presence of such type of interaction between lepton and quark tensor
currents~\cite{discovery}. If we assume the universality of the new chiral
boson coupling to lepton and quark tensor currents the value of the tensor 
coupling constant can be fixed at $g_{RR}^T\approx 0.013$,
in accordance with the explanation of the PIBETA anomaly.

Let us stress again that the most general four-fermion local
Lagrangian does not contain a tensor interaction (\ref{tensor})
and can provide no model-independent description of the  experimental
data on the muon decay. It has been shown~\cite{muon} that the most general
four-fermion Lagrangian depending on momentum transfer extends the 
local one just by two new tensor interactions with coupling
constants $g_{RR}^T$ and $g_{LL}^T$. They lead to a new parametrization
even for the case of unpolarized muon decays.

The purpose of the present paper is the calculation of 
the energy distribution and the polarization of the positron
from the decay of polarized muons in the presence of the scalar (\ref{scalar})
and the new tensor (\ref{tensor}) interactions, which
are more sensitive to experimental detection. It will be shown
how to discriminate between scalar and tensor contributions.
For example, according to Michel analysis, the large transverse 
component $P_{T_1}$ would indicate a non-zero $\eta$ parameter and
the presence of the scalar interaction (\ref{scalar}). 
On the one hand, the tensor interaction (\ref{tensor}) leads to an  
analogous contribution into $P_{T_1}$, and therefore the two contributions
cannot be distinguished in this case.
On the other hand, they lead to different energy distributions for
the isotropic part of the positron spectrum, and the anisotropic part
is affected only by the interference between the tensor and the $V-A$ 
interactions. Therefore, the combined analysis of the energy distributions and
the polarization of the positron can provide their unambiguous discrimination.

The main goal of this paper is to present the prediction of deviations from 
the SM in the energy distribution and in the polarization of the positron 
from the polarized muon decay, obtained on the assumption of the presence
of the new tensor interaction (\ref{tensor}) with the known coupling
constant $g_{RR}^T$. This analysis is very interesting today, since its
predictions could be verified in the near future by on-going experiments 
at PSI~\cite{muPT} and TRIUMF~\cite{TWIST} to a level of at least 3$\sg$.

\section{The polarized muon decay}

Taking into account the dominant contribution from the $V-A$ interaction 
(\ref{vector}) and allowing only contributions from the scalar 
(\ref{scalar}) and the tensor (\ref{tensor}) interactions, we will
calculate the differential decay probability of the polarized muon
at rest. Neglecting radiative corrections, 
the neutrino masses as well as the second and higher power of the ratio 
$x_0=2m_e/m_\mu\approx 9.7\times 10^{-3}$, its general form~\cite{review}
up to the overall normalization factor $A$ reads
\be
\frac{{\rm d}^2\Gamma}{{\rm d}x\;{\rm d}\cos\vartheta}=
\frac{m^5_\mu}{32\pi^3}\frac{A}{16}G^2_F~x
\left[F_{IS}(x)+|\mbox{\boldmath{$P$}}_\mu|
\cos\vartheta~F_{AS}(x)\right]
\left[1+\mbox{\boldmath{$\hat{\zeta}$}}\cdot
\mbox{\boldmath{$P$}}_e(x,\vartheta)\right],
\label{general}
\ee 
where $x=2E_e/m_\mu$ is the reduced energy of the positron emitted
in the direction \mbox{\boldmath{$\hat{x}$}}$_3$ at the angle $\vartheta$
with respect to the muon polarization vector \mbox{\boldmath{$P$}}$_\mu$,
and with its spin parallel to the arbitrary direction 
\mbox{\boldmath{$\hat{\zeta}$}}.
The three components of the electron polarization vector
\mbox{\boldmath{$P$}}$_e(x,\vartheta)$ are defined as
\be
\mbox{\boldmath{$P$}}_e(x,\vartheta)=P_{T_1}\mbox{\boldmath{$\hat{x}$}}_1
+P_{T_2}\mbox{\boldmath{$\hat{x}$}}_2
+P_L\mbox{\boldmath{$\hat{x}$}}_3,
\ee
where \mbox{\boldmath{$\hat{x}$}}$_1$, \mbox{\boldmath{$\hat{x}$}}$_2$
and \mbox{\boldmath{$\hat{x}$}}$_3$ are orthogonal unit vectors defined as
follows:
\be
\frac{\mbox{\boldmath{$\hat{x}$}}_3\times\mbox{\boldmath{$P$}}_\mu}
{|\mbox{\boldmath{$\hat{x}$}}_3\times\mbox{\boldmath{$P$}}_\mu|}=
\mbox{\boldmath{$\hat{x}$}}_2,\hspace{2cm}
\mbox{\boldmath{$\hat{x}$}}_2\times\mbox{\boldmath{$\hat{x}$}}_3=
\mbox{\boldmath{$\hat{x}$}}_1.
\ee

The normalization factor
\be
A=16\left(1+\frac{1}{4}|g_{RR}^S|^2+3|g_{RR}^T|^2\right)
\label{A}
\ee
is the sum of the relative probabilities for a muon to decay 
into a positron by the corresponding interactions.
%includes the squares of the absolute values of all coupling constants
%that result from the corresponding matrix elements $|{\cal M}|^2$. 
It is often forgotten or not stated explicitly that the addition of any
new interactions leads effectively to a redefinition of the Fermi
coupling constant $G_F$ for the pure $V-A$ interaction.

We will assume that the couplings constants $g_{RR}^S$ and $g_{RR}^T$ are real.
It will mean that $CP$ invariance of the interactions holds and the transverse 
component $P_{T_2}$, which is perpendicular to the plane spanned by muon 
spin and positron momentum, is zero.

In the presence of the new tensor interaction (\ref{tensor})
the isotropic $F_{IS}(x)$ and the anisotropic $F_{AS}(x)$ parts of 
the spectrum have almost standard forms:
\begin{eqnarray}
F_{IS}(x)&=&
x(1-x)+\frac{2}{9}\;\rho\;(4x^2-3x)+\eta\;x_0(1-x)+
\kappa\;x_0,
\label{IS}\\
F_{AS}(x)&=&
\frac{1}{3}\xi\;x\left[1-x+\frac{2}{3}\delta(4x-3)\right]
+\kappa\;x_0(2-x),
\label{AS}
\end{eqnarray}
where the parameters $\rho$, $\xi$ and $\delta$: 
\begin{eqnarray}
\rho&=&\frac{3}{4}
\left\{1+|g_{RR}^-|^2\right\}\frac{16}{A},
\label{rho}\\
\xi&=&\left\{1-\frac{7}{4}|g_{RR}^-|^2+\frac{3}{4}|g_{RR}^+|^2
\right\}\frac{16}{A},
\label{xi}\\
\xi\delta&=&\frac{3}{4}
\left\{1-|g_{RR}^-|^2\right\}\frac{16}{A},
\label{delta}
\end{eqnarray}
can be expressed through only two linear combinations of the coupling 
constants $g_{RR}^S$ and
$g_{RR}^T$: $g_{RR}^-=\frac{1}{2}g_{RR}^S-g_{RR}^T$ and 
$g_{RR}^+=\frac{1}{2}g_{RR}^S+3g_{RR}^T$.

The so-called ``low energy spectral shape parameter'' 
\be
\eta=\frac{1}{2}g_{RR}^S~\frac{16}{A}
\label{eta}
\ee
%is proportional to the scalar coupling constant $g_{RR}^S$ and 
contributes to the isotropic part of the spectrum (\ref{IS})
and does not affect the anisotropic part (\ref{AS}). 
It appears as a result of the interference of the vector (\ref{vector}) 
and the scalar (\ref{scalar}) interactions. 

The main feature of the new tensor interaction (\ref{tensor}) is that
the analogous parameter 
\be
\kappa=g_{RR}^T~\frac{16}{A}
\label{epsilon}
\ee
also appears as a result of the interference of the vector (\ref{vector}) 
and the tensor (\ref{tensor}) interactions and contributes to both
the isotropic (\ref{IS}) and the anisotropic (\ref{AS}) parts of the
spectrum.

The present accuracy in the experimental value of $\eta$ leads to 
an uncertainty in the value of the Fermi coupling constant $G_F$ 20 times 
larger than that of the more precisely known muon lifetime~\cite{review}.
Moreover, the new parameter $\kappa$ has never been taken into account.
Therefore, its measurement is urgently needed for a model-independent 
determination of the $G_F$. 

The non-zero components of the positron polarization vector 
\mbox{\boldmath{$P$}}$_e$ are given by
\begin{eqnarray}
P_{T_1}(x,\vartheta)&=&
{P_\mu\sin\vartheta\;F_{T_1}(x)\over
F_{IS}(x)+P_\mu\cos\vartheta~F_{AS}(x)},
\label{PT1}\\
P_{L}(x,\vartheta)&=&
{F_{IP}(x)+P_\mu\cos\vartheta~F_{AP}(x)\over
F_{IS}(x)+P_\mu\cos\vartheta~F_{AS}(x)},
\label{PL}
\end{eqnarray}
where we have used $P_\mu=|\mbox{\boldmath{$P$}}_\mu|$, and where
\be
F_{IP}(x)=\xi'\;x(1-x)+\frac{2}{9}\;\rho_L\;(4x^2-3x)
+\kappa\;x_0 x\mbox{\hspace{3cm}}
\label{IP}
\ee
and
\be
F_{AP}(x)=
\frac{1}{3}\xi''\;x\left[1-x+\frac{2}{3}\delta_L(4x-3)\right]
-\eta\;\frac{x_0}{3}(1-x)+\kappa\;\frac{x_0}{3}(1+2x)
\label{AP}
\ee
can been written down in a way analogous to $F_{IS}(x)$ and $F_{AS}(x)$
and parametrized by the parameters $\xi'$, $\rho_L$, $\xi''$ and $\delta_L$:
\begin{eqnarray}
\xi'&=&\left\{1-\frac{3}{4}|g_{RR}^-|^2-\frac{1}{4}|g_{RR}^+|^2\right\}
~\frac{16}{A},
\label{AL}\\
\rho_L&=&\frac{3}{4}
\left\{1-|g_{RR}^-|^2\right\}\frac{16}{A},
\label{rhoL}\\
\xi''&=&\left\{1+\frac{7}{4}|g_{RR}^-|^2-\frac{3}{4}|g_{RR}^+|^2
\right\}\frac{16}{A},
\label{xiL}\\
\xi''\delta_L&=&\frac{3}{4}
\left\{1+|g_{RR}^-|^2\right\}\frac{16}{A}.
\label{deltaL}
\end{eqnarray}
The transverse function $F_{T_1}(x)$ has the form
\be
F_{T_1}(x)=-\frac{x_0}{6}(1-x)\frac{16}{A}-\eta\frac{x}{3}
+\kappa\frac{x}{3},
\label{FT1}
\ee
where the negligibly small ${\cal O}(x_0|g_{RR}^\pm|^2)$ terms are
omitted.

The first term in (\ref{FT1}) is the SM contribution, which is very small,
owing to the suppression factor $x_0$, while the subsequent terms
represent scalar and tensor contributions, which are proportional 
to the new coupling constants $g_{RR}^S$ and $g_{RR}^T$ without suppression.
It is interesting to note that the measurement of only the transverse 
polarization component $P_{T_1}$ cannot allow a discrimination between
these two new contributions.

\section{Tensor interactions and right-handed neutrinos}

As far as we have no experimental indications for the presence of the 
(pseudo)scalar interactions in the weak processes and because of the
severe experimental constraint from the ratio 
$\Gamma(\pi\to e\nu)/\Gamma(\pi\to\mu\nu)$~\cite{R}, we will not consider
them in the following. The previous consideration including the scalar 
interaction (\ref{scalar}) has been done just for the illustrative purpose
of comparing its effect with that from the new tensor interaction 
(\ref{tensor}). For example, keeping only the $\eta$ parameter and omitting 
the new term with $\kappa$ could lead to a wrong fit for the isotropic
energy distribution. Later we will consider just the opposite case of  
non-zero value for $\kappa$ and $\eta=0$. We will also show that 
for the anisotropic part of the positron spectrum an effect of
the new parameter $\kappa$ may be important and
should be taken into account.  

The new tensor interaction between quark and lepton currents~\cite{MPL}
\begin{eqnarray}
{\cal L}_T^{q\ell}\hspace{-0.5cm}&&=-\sqrt{2}f_T G_F\;
\left(\bar{d}\sigma_{\alpha\lambda}u\right)
\frac{4q_\alpha q_\beta}{q^2}
\left(\bar{\nu}_e\sigma_{\beta\lambda}e_R\right)+{\rm h.c.}
\nonumber\\
&&\equiv
-\sqrt{2}f_T G\;\left(\bar{d}_R\sigma_{\alpha\lambda}u\right)
\frac{4q_\alpha q_\beta}{q^2}
\left(\bar{\nu}_e\sigma_{\beta\lambda}e_R\right)
-\sqrt{2}f_T G\;\left(\bar{d}_L\sigma_{\beta\lambda}u\right)
\left(\bar{\nu}_e\sigma_{\beta\lambda}e_R\right)+{\rm h.c.}
\label{ql} 
\end{eqnarray}
with the coupling constant $f_T\approx 0.013$ is very 
suitable~\cite{discovery} for explaining the anomaly in the branching ratio 
and in the energy distribution for the radiative pion 
decay~\cite{ISTRA,PIBETA}.
Its destructive interference with the inner bremsstrahlung process
in the radiative pion decay leads to an explanation of the lack of events
in the high-$E_\gamma$/low-$E_e$ kinematic region, where a
deficit of events was observed. It is interesting to note that
the last term in (\ref{ql}) can be rewritten in this form owing to
the identity
\be
\left(\bar{d}_L\sigma_{\alpha\lambda}u\right)
\frac{4q_\alpha q_\beta}{q^2}
\left(\bar{\nu}_e\sigma_{\beta\lambda}e_R\right)
\equiv \left(\bar{d}_L\sigma_{\beta\ld}u\right)
\left(\bar{\nu}_e\sigma_{\beta\ld}e_R\right).
\label{identity}
\ee
The equality of the coupling constants in the two last terms in (\ref{ql})
is accepted in order to avoid constraints from ordinary pion 
decay~\cite{Voloshin}.

If one assumes flavour lepton universality, it is possible to apply
the same interaction for the $\tau$-decay. For example, 
in this case the tensor interaction gives a direct contribution to 
$\tau^-\to\nu_\tau\rho^-$ decay,
because the $\rho$-meson, besides the vector coupling constant $f_\rho$, 
has also non-zero tensor coupling constant 
$f^T_\rho\simeq f_\rho/\sqrt{2}$~\cite{Braun,Becirevic,NJL}
with the tensor quark current. It leads to 5\% excess in the $\rho$ production
in $\tau$-decay with respect to the SM, which
explains deviation from CVC prediction~\cite{tau}. 

The interaction (\ref{ql}) is the effective interaction and cannot
be applied everywhere. It shows good behaviour for any values of the square
of momentum transfers besides $q^2=0$. This unphysical pole should be
cancelled out by some additional fields and interactions
in a complete theory for
the new chiral bosons, as it happens, for example, in the spontaneously
broken gauge symmetry. The interactions (\ref{ql}) and (\ref{tensor}) 
can definitely be applied as the effective interactions for decay processes, 
where $q^2$ always has positive values. However, using 
the interaction (\ref{tensor}) it should be 
impossible to calculate the cross section for the scattering process
$\nu_\mu e^-\to\mu^-\nu_e$ (``inverse muon decay''), 
where the square of the transfer momentum can be equal to zero, without
any prescription on how to handle this pole. 

Let us note also that the interaction (\ref{ql}) involves both 
the left-handed and the right-handed quarks. 
Therefore, in general, we should introduce
also the tensor interactions involving the right-handed neutrinos.
Inasmuch as the interactions with the right-handed neutrinos do not
interfere with the ordinary weak interaction, they cannot lead to the 
destructive interference observed experimentally. Therefore,
introducing the right-handed neutrinos for quark-lepton interactions
cannot explain the anomaly in the radiative pion decay 
and this introduction seems  useless.
However, in general  we should introduce such type of interactions
\be
{\cal L}_T^{q\ell R}=
-\sqrt{2}f_T G\;\left(\bar{d}_R\sigma_{\beta\lambda}u\right)
\left(\bar{\nu}_e\sigma_{\beta\lambda}e_L\right)
-\sqrt{2}f'_T G\;\left(\bar{d}_L\sigma_{\alpha\lambda}u\right)
\frac{4q_\alpha q_\beta}{q^2}
\left(\bar{\nu}_e\sigma_{\beta\lambda}e_L\right)+{\rm h.c.}, 
\label{qlR}
\ee
where the coupling constant in the first term is predicted to be
the same~\cite{MPL} as in the interaction (\ref{ql}). 
Assuming quark--lepton universality, the ratio $f'_T/f_T=2.28\pm0.21$
has been fixed from the experimental
value of the $K_S-K_L$ mass difference~\cite{KLKS}. 
A similar value $f'_T/f_T=1+\sqrt{2}\approx 2.41$
can be predicted from pure theoretical considerations, based on 
the principle of the least energy exchanged by the new massive chiral 
bosons~\cite{JINR}.

Furthermore, assuming the quark--lepton universality of the tensor interaction
we accept that the coupling constant $g_{RR}^T$ for the pure lepton 
interaction (\ref{tensor}) should have the same value as $f_T$. 
The corresponding additional interactions for pure lepton interactions
can be cast in the form
\begin{eqnarray}
{\cal L}_T^R=\hspace{-0.5cm}&&-g_{RL}^T~\sqrt{2}G_F~
\left(\bar{\mu}_L\sg_{\al\bt}\nu_\mu\right)
\left(\bar{\nu}_e\sg_{\al\bt}e_R\right)
-g_{LR}^T~\sqrt{2}G_F~
\left(\bar{\mu}_R\sg_{\al\bt}\nu_\mu\right)
\left(\bar{\nu}_e\sg_{\al\bt}e_L\right)
\nonumber\\
&&-g_{LL}^T~\sqrt{2}G_F~
\left(\bar{\mu}_L\sg_{\al\ld}\nu_\mu\right)\frac{4q_\al q_\bt}{q^2}
\left(\bar{\nu}_e\sg_{\bt\ld}e_L\right) + {\rm h.c.}~.
\label{additional}
\end{eqnarray}
If the light right-handed neutrinos exist they can contribute 
to the muon decay. However, since the neutrinos are practically massless, 
these interactions do not interfere with the ordinary $V-A$ weak interactions 
(\ref{vector}) or with the tensor interaction (\ref{tensor}), 
or among themselves. Therefore, they cannot lead to $CP$ violation
and their contributions into the muon decay are always proportional
to the square of the absolute values of their coupling constants.

The Michel parameters for the general case, when all possible tensor
interactions (\ref{tensor}), (\ref{additional}) contribute to the muon
decay, read
\begin{eqnarray}
A&=&16\left\{1+3|g_{RR}^T|^2+3|g_{RL}^T|^2+3|g_{LR}^T|^2+3|g_{LL}^T|^2\right\},
\label{AT}\\
\rho&=&\frac{3}{4}
\left\{1+|g_{RR}^T|^2+|g_{RL}^T|^2+|g_{LR}^T|^2+|g_{LL}^T|^2\right\}
\frac{16}{A},
\label{rhoT}\\
\xi&=&\left\{1+5|g_{RR}^T|^2-5|g_{RL}^T|^2+5|g_{LR}^T|^2-5|g_{LL}^T|^2
\right\}\frac{16}{A},
\label{xiT}\\
\xi\delta&=&\frac{3}{4}
\left\{1-|g_{RR}^T|^2+|g_{RL}^T|^2-|g_{LR}^T|^2+|g_{LL}^T|^2\right\}
\frac{16}{A},
\label{deltaT}\\
\xi'&=&\left\{1-3|g_{RR}^T|^2-3|g_{RL}^T|^2+3|g_{LR}^T|^2
+3|g_{LL}^T|^2\right\}~\frac{16}{A},
\label{ALT}\\
\rho_L&=&\frac{3}{4}
\left\{1-|g_{RR}^T|^2-|g_{RL}^T|^2+|g_{LR}^T|^2+|g_{LL}^T|^2\right\}
\frac{16}{A},
\label{rhoLT}\\
\xi''&=&\left\{1-5|g_{RR}^T|^2+5|g_{RL}^T|^2+5|g_{LR}^T|^2-5|g_{LL}^T|^2
\right\}\frac{16}{A},
\label{xiLT}\\
\xi''\delta_L&=&\frac{3}{4}
\left\{1+|g_{RR}^T|^2-|g_{RL}^T|^2-|g_{LR}^T|^2+|g_{LL}^T|^2\right\}
\frac{16}{A}.
\label{deltaLT}
\end{eqnarray}
The first two terms in (\ref{additional}) are the known local Michel tensor
interactions, which can be rewritten in the same form as the last term 
in (\ref{additional}) and as (\ref{tensor}), thanks to the identity 
(\ref{identity}). This form of the interactions
allows the introduction of the two new tensor coupling constants
$g_{RR}^T$ and $g_{LL}^T$, which have been set to zero in the Michel approach.
It is worth while to note also that the new interactions lead to the same
energy distribution for the isotropic spectrum (\ref{AT}), (\ref{rhoT}) as the
ordinary tensor interactions, apart from the interference term with the 
parameter $\kappa$. 

However, if we assume the presence of the light right-handed neutrinos,
the interactions (\ref{qlR}) will contribute to the ordinary pion decay
through the electromagnetic radiative corrections~\cite{Voloshin} 
due to the different coupling constants $f_T$ and $f'_T$. This will be
in contradiction with the present experimental data for the accepted
$f_T$ value. To avoid this problem we will assume that all right-handed 
neutrinos are very heavy, in accordance to the  see-saw mechanism~\cite{seesaw}
and the latest results from the oscillation experiments. Therefore,
they do not contribute either to the pion decay or
to the muon decay. Hence, only the coupling constant $g_{RR}^T$ is
relevant to low-energy physics.

\section{Comparison with experimental data and predictions}

Based on the previous considerations, we will compare existing 
experimental data on the Michel parameters with our hypothesis 
of small admixture of the new tensor interactions~(\ref{tensor})
in the muon decay. We will show that it does not contradict 
the present experimental data; moreover, it tells us where we should
be looking for deviations from the SM.    

In this section we will also make quantitative predictions for possible
deviations from the SM, which may be detected in the on-going experiments
at the PSI~\cite{muPT} and TRIUMF~\cite{TWIST}. 
We will assume that all experimental data on the muon decay could be 
described by only one parameter $g_{RR}^T$ of known value.
 
Let us begin with a discussion of the well known Michel parameters $\rho$ 
and $\eta$ for the isotropic part of the spectrum (\ref{IS}). The first
parameter $\rho$ has not been measured since 1969~\cite{Derenzo}, whilst
the most precise result, $\rho=0.7503\pm0.0026$, dates back to 
1966~\cite{Peoples}. According to eqs. (\ref{rhoT}) and (\ref{AT}) the 
deviation from the SM value of the parameter $\rho$ can be estimated as
\be
\rho=\frac{3}{4}\left\{1+|g_{RR}^T|^2\right\}\frac{16}{A}\approx
\frac{3}{4}\left\{1-2|g_{RR}^T|^2\right\}\simeq
\frac{3}{4}\left\{1-3.4\times 10^{-4}\right\},
\label{rho_pred}
\ee 
which is an order of magnitude less than the present experimental accuracy.
Its experimental value has been derived assuming $\eta=0$, because the 
parameters $\rho$ and $\eta$ exhibit a considerable statistical correlation,
when determined from fits to a limited part of the spectrum.
Therefore, even with a precision of a few parts in 10$^{-4}$ 
for the Michel parameters, the TWIST experiment could not
detect a deviation from the SM in the $\rho$ parameter.

What about the effect of the new parameter $\kappa$ on $\rho$?
First of all its effect, as well as a possible effect of the $\eta$
parameter, is suppressed by the small parameter $x_0$.
It can be shown that when the experimental distribution is compared 
to the relative event density for the theoretical spectrum, the effect 
of the new term completely diminishes. Therefore, it is practically 
impossible to detect the new contribution in the isotropic part
of the spectrum.

However, unsuppressed contributions of the interference between $V-A$
and the new interactions could be detected from the energy distribution
of the transverse polarization component $P_{T_1}$ (\ref{PT1}), (\ref{FT1}).
At the present accuracy the measured $P_{T_1}$ distribution is 
energy-independent and consistent with zero. 
The most precise experimental results 
only put constraints on the average value of $P_{T_1}$
\be
\langle P_{T_1}\rangle=0.016\pm0.023~\cite{Burkard},
\hspace{1cm}\langle P_{T_1}\rangle=0.005\pm0.016~\cite{muPT}.
\label{PT1_res}
\ee
The last one is the preliminary result of the $\mu_{P_T}$ experiment 
at the PSI.

Combining these results we obtain $\langle P_{T_1}\rangle=0.009\pm0.013$. 
In order 
to extract an effect of new physics we need to subtract the SM contribution,
which is small but not equal to zero. Both experiments were performed 
in approximately the same kinematical region where the SM predicts 
\mbox{$\langle P^{SM}_{T_1}\rangle\simeq-0.005$}. Therefore, 
the possible effect
of new physics is estimated as $\langle P^{NP}_{T_1}\rangle=0.014\pm0.013$.
This value should be compared with our prediction 
$\langle P_{T_1}^\kappa\rangle\simeq 0.015$ obtained by
taking into account the contribution from the new tensor interaction
(\ref{tensor}) with the parameter $\kappa\approx g_{RR}^T=0.013$. This 
surprising agreement can be checked in the future at the level of 3 standard
deviations by the final result of the $\mu_{P_T}$ experiment.

If a non-zero effect of new physics will be detected in $P_{T_1}$, 
the following
question will immediately arise. Is it  an effect of the interference
with the scalar interaction (\ref{scalar}) governed by the parameter $\eta$ 
or with the new tensor interaction (\ref{tensor}) described by the new
parameter $\kappa$? To answer this question let us discuss the anisotropic
part of the spectrum (\ref{AS}), which is conventionally parametrized 
by the Michel parameters $\xi$ and $\delta$. In our case we have 
an additional contribution connected with the interference between 
the SM and new tensor interactions,
while such interference is absent in the case of the scalar interactions.
We will show that the parameter $\delta$ is very sensitive to the
presence of the additional contribution.

To extract the $\xi$ and $\delta$ parameters the following asymmetry
function 
\be
{\cal A}(x)=P_\mu~\frac{F_{AS}(x)}{F_{IS}(x)}
\label{asymmetry}
\ee
is measured in the polarized muon decay. A deviation from the SM
in the parameter $\delta$ eqs.~(\ref{deltaT}), (\ref{xiT})
\be
\delta=\frac{3}{4}\left\{1+\Delta\delta\right\}
=\frac{3}{4}~\frac{1-|g_{RR}^T|^2}{1+5|g_{RR}^T|^2}
\approx\frac{3}{4}\left\{1-6|g_{RR}^T|^2\right\}
\simeq\frac{3}{4}\left\{1-10.1\times 10^{-4}\right\}
\label{deltapred}
\ee
and a non-zero value of the parameter $\kappa$ can been established 
from the precise measurement of the zero point $x_z=1/2+\Delta x_z$ 
of the asymmetry, for which ${\cal A}(x_z)=0$,
independently of the parameter $\xi$ and the polarization
$P_\mu$. It is interesting to note that both
effects act in the same direction, leading to a noticeable deviation
in $x_z$ from the SM value:
\be
\Delta x_z\approx\frac{1}{2}\Delta\delta-9\kappa x_0
\simeq-16.4\times 10^{-4}.
\label{xz}
\ee
It was at the edge of the accuracy for the previous experiment at 
TRIUMF~\cite{TRIUMF87} to detect this effect, while it could be 
pinned down at the 8$\sigma$ level in the TWIST experiment, 
thanks to the higher absolute energy calibration 10$^{-4}$ at $x=1$.

Since the new tensor interaction (\ref{ql}) does not contribute
to the pion decay \mbox{$\pi^+\to\mu^+\nu_\mu$}, the muon polarization $P_\mu$
is equal to its SM value. The deviation from the SM in the parameter $\xi$
\be
\xi=\left\{1+5|g_{RR}^T|^2\right\}\frac{16}{A}\approx 1+2|g_{RR}^T|^2
\simeq 1+3.4\times 10^{-4}
\ee
has the same value and opposite sign as in the parameter $\rho$. This value 
is too small to be detected even in the high-precision TWIST experiment
fitting the measured asymmetries to eq. (\ref{asymmetry}). At the same time,
the effect from the terms with the parameters $\delta$ and $\kappa$ should be
perceptible because of the predicted big deviation in $\delta$ 
(\ref{deltapred}) and of the large coefficient at the additional $\kappa$-term 
as in eq. (\ref{xz}). 
%Let us investigate the effect of $\kappa$-term on a determination of the
%parameter $\delta$.
For example, the previous TRIUMF experiment~\cite{TRIUMF87} had found a 
1.6$\sg$ effect in $\delta$ with the right sign as predicted
in this paper; the authors had nevertheless assigned this purely to statistics.

The parameter $\xi$ can be measured also from the integral asymmetry
\be
{\cal A}'=P_\mu
~\frac{\int F_{AS}(x){\rm d}x}{\int F_{IS}(x){\rm d}x}.
\ee
Although the parameters $\rho$ and $\delta$ do not contribute to this ratio
and the predicted deviation in the parameter $\xi$ is small, the interference
$\kappa$-terms lead to the dominant contribution
\be
{\cal A}'\approx
\frac{1}{3}P_\mu\xi\frac{1+24\kappa x_0}{1+6\kappa x_0}
\approx\frac{1}{3}P_\mu
\left\{1+18x_0 g_{RR}^T+2|g_{RR}^T|^2\right\}
\simeq\frac{1}{3}P_\mu\left\{1+26.0\times 10^{-4}\right\}.
\ee
The predicted value does not contradict  the best present measurements:
\be 
P_\mu\xi=1.0027\pm0.0079\pm0.0030~\cite{Beltrami},\hspace{1cm} 
P_\mu\xi=1.0013\pm0.0030\pm0.0053~\cite{Imazato},
\ee
and can be verified in the TWIST experiment from a slope determination 
of the $\cos\vartheta$ distribution.

To conclude our analysis let us consider the effect of the new tensor 
interaction
on the total decay rate $\Gamma$ and the $G_F$ determination, and how it is
connected to the unitarity problem for the first row of the CKM matrix.
As has been shown in ref.~\cite{muon} and as follows also from 
eq.~(\ref{general}) at $\eta=0$, the total decay rate is
\be
\Gamma=\frac{m_\mu^5 G_F^2}{192\pi^3}
\left\{1+6x_0 g_{RR}^T+3|g_{RR}^T|^2\right\}.
\ee
Then in the presence of the new tensor interaction the experimental value 
$G_F^{\rm exp}=(1.16639\pm 0.00001)\times 10^{-5}$ GeV$^{-2}$
derived from the muon lifetime is related to the Fermi coupling constant 
$G_F$ as
\be
G_F^{\rm exp}=\sqrt{1+6x_0 g_{RR}^T+3|g_{RR}^T|^2}~G_F
\simeq 1.00063 G_F.
\ee
Therefore, the real $G_F$ value could be half per mille lower than usually
accepted. The matrix element $V_{ud}^{\rm exp}=0.9740\pm 0.0005$~\cite{Vud} 
is extracted from the super-allowed $0^+\to 0^+$ Fermi transitions, using 
the current experimental value of $G_F^{\rm exp}$.
Therefore, in order to obtain its real value
\be
V_{ud}=\frac{G_F^{\rm exp}}{G_F}V_{ud}^{\rm exp}\simeq 0.9746\pm 0.0005,
\label{Vud}
\ee
the corrections in $G_F$ have been applied. This value can be accepted
as real one, because the new tensor interaction (\ref{ql}) contributes only
to the Gamow--Teller transitions and do not affect $V_{ud}^{\rm exp}$
determination. 

Using this fact we can calculate the matrix element $V_{us}$ from 
the unitarity condition 
\be
V_{us}=\sqrt{1-|V_{ud}|^2-|V_{ub}|^2}=0.2239\pm 0.0022,
\label{Vus}
\ee
where the contribution from $V_{ub}$ can safely be neglected with 
respect to the accuracy to which $V_{ud}$ can be determined.
This value surprisingly coincides with the value 0.2238(30)~\cite{ratio}
determined from the ratio of experimental kaon and pion decay widths
$\Gamma(K\to\mu\nu)/$ $\Gamma(\pi\to\mu\nu)$~\cite{PDG} using
the lattice calculations of the pseudoscalar decay constant 
ratio $f_K/f_\pi$~\cite{lattice} and assuming unitarity.
Here it should be noted that the new tensor interactions do not contribute
to the two-body decay of the pseudoscalar mesons. Therefore, the previous
determination of $V_{us}$ is valid also in the presence of the new 
interactions.
This fact confirms the recent result of the E865 Collaboration on the
$K_{e3}^+$ branching ratio~\cite{E865} since the extracted $V_{us}$ value
0.2238(33)~\cite{Ke3} is in good agreement with unitarity and 
the $V_{ud}$ determination from nuclear super-allowed beta decays. 
Since the value of $V_{us}$ is small, even if the new tensor interactions
have an effect on the kaon decays, a deviation in this value will be
one order of magnitude smaller than its accuracy.

\section*{Acknowledgements}
I would like to thank A. Arbuzov, A. Gaponenko, P. Gorbunov, W. Fetscher,
D. Kirilova, M.~Mateev, V. Obraztsov, W. Sobk\'ow for their interest 
in this work and their overall help. 
I acknowledge the warm hospitality of the Theory Division at CERN,
where this work has been done.

%\pagebreak[2]

\end{document}